\documentclass[conference]{IEEEtran}
\IEEEoverridecommandlockouts
\usepackage{cite}
\usepackage{amsmath,amssymb,amsfonts}
\usepackage{algorithmic}
\usepackage[ruled]{algorithm2e}
\usepackage{graphicx}
\usepackage{textcomp}
\usepackage{xcolor}
\def\BibTeX{{\rm B\kern-.05em{\sc i\kern-.025em b}\kern-.08em
    T\kern-.1667em\lower.7ex\hbox{E}\kern-.125emX}}
\begin{document}

\title{A Knowledge-Driven Meta-Learning Method for CSI Feedback\\
}
\author{\IEEEauthorblockN{Han Xiao\textsuperscript{1}, Wenqiang Tian\textsuperscript{1}, Wendong Liu\textsuperscript{1}, Zhi Zhang\textsuperscript{1}, Zhihua Shi\textsuperscript{1}, Li Guo\textsuperscript{1} and Jia Shen\textsuperscript{1}}
\IEEEauthorblockA{\textsuperscript{1}Department of Standardization, OPPO Research Institute, Beijing, China \\
Email: \{xiaohan1, tianwenqiang, liuwendong1, zhangzhi, szh, v-guoli, sj\}@oppo.com}

}

\maketitle

\begin{abstract}
Accurate and effective channel state information
(CSI) feedback is a key technology for massive multiple-input and
multiple-output (MIMO) systems. Recently, deep learning (DL) has been introduced to enhance CSI feedback in massive MIMO application, where the massive collected training data and lengthy training time are costly and impractical for realistic deployment. In this paper, a knowledge-driven meta-learning solution for CSI feedback is proposed, where the DL model initialized by the meta model obtained from meta training phase is able to achieve rapid convergence when facing a new scenario during the target retraining phase. Specifically, instead of training with massive data collected from various scenarios, the meta task environment is constructed based on the intrinsic knowledge of spatial-frequency characteristics of CSI for meta training. Moreover, the target task dataset is also augmented by exploiting the knowledge of statistical characteristics of channel, so that the DL model initialized by meta training can rapidly fit into a new target scenario with higher performance using only a few actually collected data in the target retraining phase. The method greatly reduces the demand for the number of actual collected data, as well as the cost of training time for realistic deployment. Simulation results demonstrate the superiority of the proposed approach from the perspective of feedback performance and convergence speed.
\end{abstract}


\section{Introduction}
Accurate and effective channel state information
(CSI) feedback has been intensively studied for supporting massive multiple-input and multiple-output (MIMO) systems. Along with the standardization in the 3rd Generation Partnership Project (3GPP), various solutions based on the TypeI and enhanced TypeII (eTypeII) codebook have been proposed to improve the CSI feedback performance \cite{TS0002}. However, to resolve the issues of larger feedback overhead and insufficient recovery accuracy, methods for further enhancing the CSI feedback are still being actively studied.  

Recently, deep learning (DL) has been introduced for CSI feedback enhancement, where the DL model can achieve higher CSI recovery accuracy with reduced feedback overhead. An autoencoder method of CsiNet for CSI feedback \cite{wen2018deep} is first proposed, where an encoder at the user equipment (UE) compresses the channel matrix and a decoder at the base station (BS) recovers the corresponding channel matrix. Subsequently, a series of follow-up works are conducted under various conditions \cite{li2020spatio,xiao2022ai,guo2021canet,Xiao2021AIEW,liu2021evcsinet} . However, there are still some challenges for DL-based CSI feedback. First, the generalization issue should be considered since the DL methods tend to express the scenario-specific property. Moreover, plenty of training data of target scenario is quite impractical for deployment due to the expense and long-time training and collecting data. Meta-learning is utilized for CSI feedback in \cite{zeng2021downlink} and \cite{tolba2021meta}, where the model is initialized by the meta model obtained in meta training phase with massive CSI samples corresponding to multiple various scenarios, and then achieves quick convergence with small amount of CSI data in a new target scenario. However, the above meta-learning based solutions still require massive collected data for the meta training phase. Moreover, in target retraining phase the model is retrained on the original small amount of data within short time, thus it might suffer from performance loss in the new target scenario in comparison with models trained on sufficient data. Further, the above methods also fail to consider the knowledge of intrinsic characteristics of the wireless communication during both phases.

In this paper, a novel knowledge-driven meta-learning method for CSI feedback is proposed. Specifically, instead of training with massive CSI data collected from different wireless scenarios in meta training phase, one can construct the meta task environment by exploring the intrinsic knowledge of spatial-frequency characteristic of CSI eigenvector for meta training. After the DL model obtains the initialization in meta training phase, it is capable of achieving rapid convergence by retraining on target task dataset, which is augmented from only small amount of actually collected seeded data with the assistance of the knowledge of statistical feature of wireless channels. Simulation results illustrate the superiority of the proposed method from the perspective of feedback performance and convergence speed. 

{\it Notations}: uppercase and lowercase letters denote scalars. Boldface uppercase and boldface lowercase letters denote matrices and vectors, respectively. Calligraphic uppercase letters denote sets. $\mathbf{A}(:,\mathcal{B})$ and $\mathbf{A}(\mathcal{B},:)$ denote the sub-matrices of $\mathbf{A}$ that consist of the columns and rows indexed by set $\mathcal{B}$, respectively. $\mathbb{E}\{\cdot\}$ denotes expectation and $\mathrm{Tr}\{\cdot\}$ denotes trace. $\mathbf{A}^{\rm H}$ denotes the Hermitian matrix of $\mathbf{A}$. $\mathrm{rand}(\mathcal{A}, a)$ denotes the random sampling of $a$ samples from set $\mathcal{A}$ without replacement. The sets of real and complex numbers are denoted by $\mathbb{R}$ and $\mathbb{C}$, respectively. $|\cdot|$ denotes the cardinality of a set or the absolute value of a scalar.

\section{System Description}\label{System Description}
\subsection{System Model}
A MIMO system with $N_{\rm{t}} = N_{\rm h}N_{\rm v}$ transmitting antennas at BS and $N_{\rm{r}}$ receiving antennas at UE is considered, where $N_{\rm h}$ and $N_{\rm v}$ are the numbers of horizontal and vertical antenna ports, respectively.
Note that our proposed methods are suitable for antennas with either dual or single polarization, and that single polarization is considered to illustrate the basic principle in this paper. The downlink channel in time domain can be denoted as a three-dimensional matrix $\widehat{\mathbf{H}} \in \mathbb{C}^{N_{\rm r}\times N_{\rm t}\times N_{\rm d}}$, where $N_{\rm d}$ is the number of paths with various delays. By conducting Discrete Fourier transform (DFT) over the delay-dimension of the time-domain downlink channel matrix $\widehat{\mathbf{H}}$, the downlink channel in frequency domain $\widetilde{\mathbf{H}}\in\mathbb{C}^{N_{\rm r}\times N_{\rm t}\times N_{\rm sc}}$ can be written as
\begin{equation}\label{Hf_1}
\widetilde{\mathbf{H}}= \big[\widetilde{\mathbf{H}}_1,\widetilde{\mathbf{H}}_2,\cdots,\widetilde{\mathbf{H}}_{N_{\rm{sc}}}\big],
\end{equation}
where $N_{\rm{sc}}$ is the number of subcarriers, and $\mathbf{H}_k\in\mathbb{C}^{N_{\rm r}\times N_{\rm t}}, 1\leq k\leq N_{\rm sc}$ denotes the downlink channel on the $k$th subcarrier. Normally, the CSI eigenvector feedback is performed on each subband which consists of $N_{\rm gran}$ subcarriers with $N_{\rm sc} = N_{\rm gran}N_{\rm sb}$. Assuming the rank 1 configuration for downlink transmission, the corresponding eigenvector for the $l$th subband $\mathbf{w}_l\in\mathbb{C}^{N_{\rm t}\times 1}$ with $||\mathbf{w}_l||_2=1$, can be calculated by the eigenvector decomposition on the subband as
\begin{equation}
\left(\frac{1}{N_{\rm{gran}}}\sum_{k=(l-1)N_{\rm{gran}}+1}^{lN_{\rm{gran}}}\widetilde{\mathbf{H}}_k^{\rm H}\widetilde{\mathbf{H}}_k\right) \mathbf{w}_l =\lambda_l \mathbf{w}_l,
\end{equation}
where $1\leq l \leq N_{\rm sb}$ and $\lambda_l$ represents the corresponding maximum eigenvalue for the $l$-th subband. Therefore,  the CSI eigenvector for all $N_{\rm{sb}}$ subbands can be written as
\begin{equation}\label{Wi}
\mathbf{W}= \big[\mathbf{w}_{1},\mathbf{w}_{2},\cdots,\mathbf{w}_{N_{\rm{sb}}}\big] \in \mathbb{C}^{N_{\rm t}\times N_{\rm{sb}}},
\end{equation}
wherein total $N_{\rm{sb}}N_{\rm t}$ complex coefficients need to be compressed at the UE and then recovered at the BS side.

Generally, the optimization objective for CSI feedback can be given as
\begin{equation}\label{score_function_1}
\begin{split}
\min_{\mathfrak{F}} -\rho(\mathbf{W}, \mathbf{W}')
 = \min_{\mathfrak{F}} -\frac{1}{N_\textrm{sb}}\sum_{l=1}^{N_\textrm{sb}}\left(\frac{\|\mathbf{w}^{\rm H}\mathbf{w}'\|_2}{\|\mathbf{w}\|_2\|\mathbf{w}'\|_2}\right)^2
\end{split}
\text{,}
\end{equation}
where $\rho(\cdot,\cdot) \in [0,1]$ denotes the squared generalized cosine similarity (SGCS), $\|\cdot\|_2$ denotes $\ell_{2}$ norm, $\mathbf{w}_{l}$ and $\mathbf{w}'_{l}$ represent the original and recovered CSI eigenvector of the $l$-th subband, respectively, $\mathfrak{F}$ represents the alternative CSI feedback schemes such as TypeI, eTypeII and DL-based autoencoder.

\subsection{DL-based CSI Feedback}
\begin{figure}[tb]
\setlength{\abovecaptionskip}{-0.03cm}
\setlength{\belowcaptionskip}{-1cm}
\centering
\includegraphics[scale=0.9]{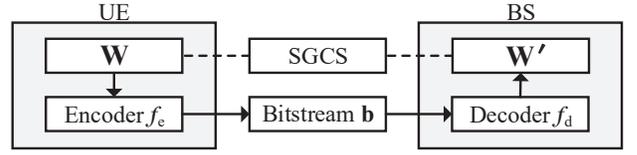}
\caption{Illustration of DL-based CSI feedback.}
\label{basic_autoencoder}
\vspace{-0.5cm}
\end{figure}

The architecture of DL-based CSI feedback using autoencoder is introduced in Fig. \ref{basic_autoencoder}, where the neural network (NN) encoder and decoder, $f_{\rm e}(\cdot;\Theta_{\rm{E}})$ and $f_{\rm d}(\cdot;\Theta_{\rm{D}})$ with trainable parameters $\Theta = \{\Theta_{\rm{E}}, \Theta_{\rm{D}}\}$ are deployed at UE and BS, respectively. Thus the DL-based autoencoder $f_{\rm a}(\cdot; \Theta)$ with trainable parameters $\Theta = \{\Theta_{\rm{E}}, \Theta_{\rm{D}}\}$ can be represented as
\begin{equation}\label{autoencoder_1}
\begin{split}
\mathbf{W}' =  f_{\rm d}(f_{\rm e}(\mathbf{W};\Theta_{\rm{E}}); \Theta_{\rm{D}}) = f_{\rm a}(\mathbf{W};\Theta)
\end{split}
\text{,}
\end{equation}
where the encoder first compresses and quantizes the original CSI eigenvector $\mathbf{W}$ to a bitstream $\mathbf{b}$ of length $B$. Then the decoder uses $\mathbf{b}$ to recover $\mathbf{W}'$. During training phase, the encoder and decoder are jointly optimized to solve (\ref{score_function_1}) with sufficient numbers of CSI eigenvector samples.

\subsection{Meta-learning based CSI Feedback}
Generally, the goal of meta-learning based CSI feedback is to find a good initialization of $\Theta = \{\Theta_{\rm{E}}, \Theta_{\rm{D}}\}$, so that the autoencoder can converge quickly with a small amount of CSI samples and a few training steps for a new scenario. Specifically, the procedure of meta-learning based CSI feedback can be divided into two phases, i.e., the meta training phase and target retraining phase.

During meta training phase, the model is trained over a big dataset consisting of $T$ CSI tasks of diverse scenarios, which can be defined as meta task environment $\mathcal{T}_{\rm meta} = \{\mathcal{T}_1,...,\mathcal{T}_{T}\}$, wherein each task $\mathcal{T}_j = \{\mathbf{W}^{j}_{1},...,\mathbf{W}^{j}_{T}\}, 1\leq j \leq T$ consists of $|\mathcal{T}_j|$ CSI samples denoted as $\mathbf{W}^{j}_{i}, 1\leq i \leq \mathcal{T}_j$. Based on the meta task environment $\mathcal{T}_{\rm meta}$, meta-learning algorithms can be performed to learn the initial parameters $\widehat{\Theta}$, i.e.,
\begin{equation}\label{meta_training_pahse_problem}
\begin{split}
\min_{\widehat{\Theta}} \mathbb{E}_{\mathcal{T}_j \subset \mathcal{T}_{\rm meta}}\left[-\rho'(\mathcal{T}_j,f_{\rm a}(\mathcal{T}_j;\mathrm{U}^g_{\mathcal{T}_j}(\widehat{\Theta})))   \right]
\end{split}
\text{,}
\end{equation}
where $\mathrm{U}^g_{\mathcal{T}_j}(\widehat{\Theta})$ is the operator that updates $\widehat{\Theta}$ for $g$ training steps using data sampled from $\mathcal{T}_j$. The initialization $\widehat{\Theta}$ learnt in (\ref{meta_training_pahse_problem}) is expected to has the same ability of quick adaptation with small amount of data on an unobserved target task $\mathcal{T}_{\rm target}$.

\begin{figure*}[t]

\setlength{\belowcaptionskip}{-1cm}
\centering
\includegraphics[scale=0.72]{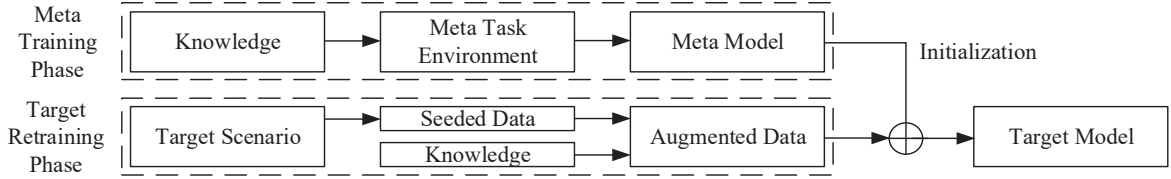}
\caption{Proposed knowledge-driven meta-learning framework for CSI feedback.}
\label{pic_proposed}
\vspace{-0.4cm}
\end{figure*}

Secondly, the target retraining phase can be formulated as
\begin{equation}\label{target_retraining_pahse_problem}
\begin{split}
\min_{\Phi = \mathrm{U}^g_{\mathcal{T}_{\rm target}}(\widehat{\Theta})} -\rho'(\mathcal{T}_{\rm target},f_{\rm a}(\mathcal{T}_{\rm target};\Phi))
\end{split}
\text{,}
\end{equation}
where $\Phi$ denotes the possible parameter sets trained on $\mathcal{T}_{\rm target}$ after $g$ retraining steps based on the initialization $\widehat{\Theta}$, which indicates that the final parameters on a new target task of scenario can be rapidly obtained with only a few retraining steps.

However, the existing meta-leaning based CSI feedback still has to face two major challenges, which our knowledge-driven meta-learning method aims to solve.
\begin{itemize}
\item During meta training phase, it requires sufficient samples to construct the meta task environment $\mathcal{T}_{\rm meta}$ to solve (\ref{meta_training_pahse_problem}), which is extremely costly since it is impractical to collect all existing types of wireless scenarios with adequate diversity.
\item During target retraining phase, despite the rapid convergence for solving (\ref{target_retraining_pahse_problem}) using small amount of data $\mathcal{T}_{\rm target}$ based on the initialization $\widehat{\Theta}$, it is always difficult to achieve comparable performance with using large amount of CSI data in target scenario.
\end{itemize}

\section{Knowledge-driven Meta-Learning for CSI Feedback}\label{Knowledge-driven Meta-Learning for CSI Feedback}
\subsection{Knowledge-driven Meta Training Phase}
\subsubsection{Spatial-Frequency Characteristic}\label{Spatial-Frequency Characteristic}
Generally, considering the intrinsic structure of the CSI eigenvector, $\mathbf{W} \in \mathbb{C}^{N_{\rm t}\times N_{\rm{sb}}}$ can be decomposed as
\begin{equation}\label{decompose}
\setlength{\abovedisplayskip}{3pt}
\setlength{\belowdisplayskip}{3pt}
\mathbf{W} = \mathbf{S}\mathbf{E}\mathbf{F}^{\rm H}
\end{equation}
where $\mathbf{S}\in \mathbb{C}^{N_{\rm t} \times N_{\rm t}}$ is constructed with $N_{\rm t}$ orthogonal basis vectors in spatial domain and $\mathbf{F} \in \mathbb{C}^{N_{\rm sb} \times N_{\rm sb}}$ is constructed with $N_{\rm sb}$ orthogonal basis vectors in frequency domain. Specifically, both $\mathbf{S}$ and $\mathbf{F}$ are unitary matrices, which indicate the full-rank spatial-frequency characteristic. The projection coefficient matrix $\mathbf{E} \in \mathbb{C}^{N_{\rm t} \times N_{\rm sb}}$ represents that each CSI eigenvector $\mathbf{W}$ can be completely expressed by the linear combination of the orthogonal basis vectors in $\mathbf{S}$ and $\mathbf{F}$. Obviously, the distribution of the elements in $\mathbf{E}$ with relatively larger amplitude determines the dominant spatial-frequency feature of $\mathbf{W}$ given the same $\mathbf{S}$ and $\mathbf{F}$, where the dominant spatial-frequency features can be considered as the intrinsic knowledge and hence can be learnt by the DL model during the meta-training phase.

\subsubsection{Knowledge-driven Meta Training}
Inspired by the intrinsic knowledge of spatial-frequency feature in section \ref{Spatial-Frequency Characteristic}, a knowledge-driven algorithm is proposed to solve (\ref{meta_training_pahse_problem}). During the meta training phase, the meta task environment $\mathcal{T}_{\rm meta} = \{\mathcal{T}_1,...,\mathcal{T}_{T}\}$ consisting of $T$ tasks is firstly established, where the construction approach of CSI eigenvector in each task explores the CSI decomposition formula in section \ref{Spatial-Frequency Characteristic}. Each task usually consists of different CSI eigenvectors from specific number of UEs that can be sampled on various number of slots. Specifically, denote $N_{\rm ue\it, j}$ and $N_{\rm slot\it, j}$ as the number of UEs and slots for the $j$-th task $\mathcal{T}_j$, $1\leq j \leq T$, respectively, which can be set as
\begin{equation}\label{random_ue}
\setlength{\abovedisplayskip}{3pt}
\setlength{\belowdisplayskip}{3pt}
\begin{split}
N_{\rm ue,\it j} = \mathrm{rand}(\{1,...,\widehat{N}_{\rm ue}\}, 1)
\end{split}
\text{,}
\end{equation}
\begin{equation}\label{random_slot}
\setlength{\abovedisplayskip}{3pt}
\setlength{\belowdisplayskip}{3pt}
\begin{split}
N_{\rm slot,\it j} = \mathrm{rand}(\{1,...,\widehat{N}_{\rm slot}\}, 1)
\end{split}
\text{,}
\end{equation}
where $N_{\rm slot,\it j}N_{\rm ue,\it j} = |\mathcal{T}_j|$, $\widehat{N}_{\rm ue}$ and $\widehat{N}_{\rm slot}$ denote the maximum number of UEs and maximum number of slots of CSI that can be generated in one task, respectively.

Moreover, according to the intrinsic knowledge of spatial-frequency feature, to generate the CSI samples in the $j$-th task $\mathcal{T}_j$, $P$ groups of spatial orthogonal basis vector and one group of frequency orthogonal basis vector can be firstly given as
\begin{equation}\label{special_basis}
\setlength{\abovedisplayskip}{3pt}
\setlength{\belowdisplayskip}{3pt}
\begin{split}
\mathbf{S}_{p} = [\mathbf{s}_{p,1},...,\mathbf{s}_{p,N_{\rm t}}] \in \mathbb{C}^{N_{\rm t} \times N_{\rm t}}, 1\leq p \leq P
\end{split}
\text{,}
\end{equation}
\begin{equation}\label{frequency_basis}
\setlength{\abovedisplayskip}{3pt}
\setlength{\belowdisplayskip}{3pt}
\begin{split}
\mathbf{F} = [\mathbf{f}_{1},...,\mathbf{f}_{N_{\rm sb}}] \in \mathbb{C}^{N_{\rm sb} \times N_{\rm sb}}
\end{split}
\text{,}
\end{equation}
respectively, where each column of $\mathbf{S}_{p}$ and $\mathbf{F}$ is an orthogonal basis vector. Specifically, each basis in $\mathbf{S}_p$ indicates a beam direction in spatial domain, and multiple groups of orthogonal basis vectors are designed in order to improve the diversity of spatial features. Here we introduce a Schmidt orthogonalization method for obtaining the basis vector groups $\mathbf{S}_{p}$ and $\mathbf{F}$. For each group of spatial orthogonal basis vector $\mathbf{S}_p, 1\leq p\leq P$, and the frequency orthogonal basis vector group $\mathbf{F}$, the Schmidt orthogonalization can be performed on three full-rank random matrices $\mathbf{X}_{p}^{\rm h}\in \mathbb{C}^{N_{\rm h} \times N_{\rm h}}\sim \mathcal{CN}(0,1)$, $\mathbf{X}_{p}^{\rm v} \in \mathbb{C}^{N_{\rm v} \times N_{\rm v}}\sim \mathcal{CN}(0,1)$ and $\mathbf{X}^{\rm f} \in \mathbb{C}^{N_{\rm sb} \times N_{\rm sb}}\sim \mathcal{CN}(0,1)$, obtaining the orthogonal matrices $\mathbf{U}^{\rm h}_{p}$, $\mathbf{U}^{\rm v}_{p}$ and $\mathbf{U}^{\rm f}$, respectively. $\mathbf{F} = \mathbf{U}^{\rm f}$ can be utilized as frequency orthogonal basis vector. The $p$-th spatial orthogonal basis vector group can be obtained by performing kronecker product, i.e., $\mathbf{S}_p = \mathbf{U}^{\rm h}_{p} \otimes \mathbf{U}^{\rm v}_{p}$

Next, the method of generating CSI samples for the $j$-th task $\mathcal{T}_j$ is introduced. The group index $p_j$ for task $\mathcal{T}_j$ are first randomized by
\begin{equation}\label{random_group}
\setlength{\abovedisplayskip}{3pt}
\setlength{\belowdisplayskip}{3pt}
\begin{split}
\{p_j\} = \mathrm{rand}(\{1,...,P\}, 1)
\end{split}
\text{,}
\end{equation}
and the indices of dominant spatial and frequency feature vectors are also randomized by
\begin{equation}\label{random_s_0}
\setlength{\abovedisplayskip}{3pt}
\setlength{\belowdisplayskip}{3pt}
\begin{split}
\widehat{\mathcal{S}}_{j} = \mathrm{rand}(\{1,...,N_{\rm t}\}, L_{\rm task})
\end{split}
\text{,}
\end{equation}
\begin{equation}\label{random_f_0}
\begin{split}
\widehat{\mathcal{F}}_{j} = \mathrm{rand}(\{1,...,N_{\rm sb}\}, M_{\rm task})
\end{split}
\text{,}
\end{equation}
respectively, where the parameters $L_{\rm task} \leq N_{\rm t}$ and $M_{\rm task} \leq N_{\rm sb}$ are defined to constrain the degree of feature diversity of the task in spatial and frequency domain, respectively.

For the $m$-th UE $1\leq m \leq N_{\rm ue}$ in task $\mathcal{T}_j$, the indices of the dominant spatial and frequency feature vectors are also randomized by
\begin{equation}\label{random_s_1}
\setlength{\abovedisplayskip}{3pt}
\setlength{\belowdisplayskip}{3pt}
\begin{split}
\widetilde{\mathcal{S}}_{m} = \mathrm{rand}(\widehat{\mathcal{S}}_{j}, L_m)
\end{split}
\text{,}
\end{equation}
\begin{equation}\label{random_f_1}
\begin{split}
\widetilde{\mathcal{F}}_{m} = \mathrm{rand}(\widehat{\mathcal{F}}_{j}, M_m)
\end{split}
\text{,}
\end{equation}
respectively, where the degree of feature diversity in spatial and frequency domain $L_{\rm m}$ and $M_{\rm m}$ are both UE-specific, i.e.,
\begin{equation}\label{random_L_m}
\setlength{\abovedisplayskip}{3pt}
\setlength{\belowdisplayskip}{3pt}
\begin{split}
\{L_{m}\} = \mathrm{rand}(\{1,...,L_{\rm task}\}, 1)
\end{split}
\text{,}
\end{equation}
\begin{equation}\label{random_M_m}
\setlength{\abovedisplayskip}{3pt}
\setlength{\belowdisplayskip}{3pt}
\begin{split}
\{M_{m}\} = \mathrm{rand}(\{1,...,M_{\rm task}\}, 1)
\end{split}
\text{,}
\end{equation}
respectively.

\begin{algorithm}[t]
\caption{Knowledge-driven Meta Training Phase}
\label{alg_1}
\textbf{Initialization}:$\widehat{N}_{\rm ue}$, $\widehat{N}_{\rm slot}$, $T$, $\alpha$, $\beta$, $g$, $\epsilon$, $\widehat{\Theta}$\;
Formulate the feature basis using (\ref{special_basis}) to (\ref{frequency_basis})\;
\For{$j = 1,\ldots,T$}{
    Construct structure of $\mathcal{T}_{j}$ using (\ref{random_ue}), (\ref{random_slot}) and (\ref{random_group}) to (\ref{random_f_0})\;
    \For{$m = 1,\ldots,N_{\rm ue, \it j}$}{
    Consturt structure of UE $m$ using (\ref{random_s_1}) to (\ref{random_M_m})\;
    \For{$n = 1,\ldots,N_{\rm slot, \it j}$}{
    Generate a CSI of slot $n$ using (\ref{random_s_2}) to (\ref{csi_norm_1})\;
	}
	}
}
Meta training by iterating (\ref{meta_training_1}).
\end{algorithm}

Similarly for the $n$-th slot $1\leq n \leq N_{\rm slot}$ of the $m$-th UE in task $\mathcal{T}_j$, the dominant spatial and frequency feature vectors are respectively selected from the corresponding dominant vectors of the UE, so that the feature is maintained for the $m$-th UE but distinguished between different slots, i.e.,
\begin{equation}\label{random_s_2}
\begin{split}
\mathcal{S}_{m,n} = \mathrm{rand}(\widetilde{\mathcal{S}}_{m}, \lceil\alpha L_m\rceil)
\end{split}
\text{,}
\end{equation}
\begin{equation}\label{random_f_2}
\begin{split}
\mathcal{F}_{m,n} = \mathrm{rand}(\widetilde{\mathcal{F}}_{m},  \lceil \beta M_m \rceil)
\end{split}
\text{,}
\end{equation}
where the parameters $\alpha \in (0,1]$ and $\beta \in (0,1]$ are set to scale the diversity of the feature of each slot. Consequently, a CSI sample for the $n$-th slot of the $m$-th UE in task $\mathcal{T}_j$ can be generated as
\begin{equation}\label{csi_gen_1}
\begin{split}
\mathbf{W}^{j}_{m,n} = \mathbf{S}_{p_j}(:,\mathcal{S}_{m,n}) \widehat{\mathbf{E}}\mathbf{F}^{\rm H}(:,\mathcal{F}_{m,n})
\end{split}
\text{,}
\end{equation}
where the elements in $\widehat{\mathbf{E}} \in \mathbb{C}^{|\mathcal{S}_{m,n}| \times |\mathcal{F}_{m,n}|}$ are independently sampled from complex normal distribution $\mathcal{CN}(0,1)$. A subband-level normalization should be also performed for $1\leq l\leq N_{\rm sb}$ using
\begin{equation}\label{csi_norm_1}
\begin{split}
\mathbf{W}^{j}_{m,n}(:,l) = \frac{\mathbf{W}^{j}_{m,n}(:,l)}{||\mathbf{W}^{j}_{m,n}(:,l)||_2}
\end{split}
\text{.}
\end{equation}
Through the procedure of (\ref{random_ue}) to (\ref{csi_norm_1}) for generating each CSI sample of each UE, the meta task environment $\mathcal{T}_{\rm meta}$ can be finally constructed.

Utilizing the meta task environment $\mathcal{T}_{\rm meta}$, the meta training procedure can be conducted to solve (\ref{meta_training_pahse_problem}). The parameters of the DL model of CSI feedback is randomly initialized by $\widehat{\Theta}$. For the $j$-th task $\mathcal{T}_j$ in the meta task environment $\mathcal{T}_{\rm meta}$, $\widehat{\Theta}$ can be updated with
\begin{equation}\label{meta_training_1}
\begin{split}
\widehat{\Theta} = \widehat{\Theta} + \epsilon (\mathrm{U}^g_{\mathcal{T}_j}(\widehat{\Theta}) - \widehat{\Theta})
\end{split}
\text{,}
\end{equation}
where $\mathrm{U}^g_{\mathcal{T}_j}(\widehat{\Theta})$ is the operator that updates $\widehat{\Theta}$ for $g$ training steps on task $\mathcal{T}_j$, and $\epsilon$ denotes the step size of meta training. After that, the obtained $\widehat{\Theta}$ can be utilized as initialization for further fast retraining on a new target task of scenario. The proposed algorithm for knowledge-driven meta training phase is summarized in Algorithm \ref{alg_1}.

\subsection{Knowledge-driven Target Retraining Phase}
\subsubsection{Statistical Feature of Channel}\label{Statistical Feature of Channel}
In this part, the statistical features of the channel in both spatial domain and time delay domain are explored. Specifically, for a specific UE in the target scenario, denote the actually collected $\widetilde{N}_{\rm slot}$ channel samples in time domain as $\mathcal{H} = \{\widehat{\mathbf{H}}_1,...,\widehat{\mathbf{H}}_{\widetilde{N}_{\rm slot}}\}$, where each channel sample $\widehat{\mathbf{H}}_t \in \mathbb{C}^{N_{\rm r}\times N_{\rm t}\times N_{\rm d}}$, $1\leq t \leq \widetilde{N}_{\rm slot}$.

Firstly, the statistical feature in delay domain can be described by the power-delay spectrum. Denote $\widehat{\mathbf{H}}'_{t,d} \in \mathbb{C}^{N_{\rm r}\times N_{\rm t}}, 1\leq d \leq N_{\rm d}$ as the $d$-th delay of the $t$-th channel sample $\widehat{\mathbf{H}}_t$, the power of the $d$-th delay can be calculated as
\begin{equation}\label{cal_DPS}
\setlength{\abovedisplayskip}{3pt}
\setlength{\belowdisplayskip}{3pt}
\begin{split}
\hat{p}_d = \frac{1}{N_{\rm t}N_{\rm r}\widetilde{N}_{\rm slot}}\sum_{t=1}^{\widetilde{N}_{\rm slot}}||\widehat{\mathbf{H}}'_{t,d}||_{\rm F}^2
\end{split}
\text{,}
\end{equation}

Secondly, the statistical feature in spatial domain can be demonstrated by the self-correlation matrices of the transmitting and receiving antenna ports, which can be calculated as
\begin{equation}\label{R_tx}
\setlength{\abovedisplayskip}{3pt}
\setlength{\belowdisplayskip}{3pt}
\begin{split}
\mathbf{R}^{\rm tx}_{d} = \frac{N_{\rm t}\sum_{t=1}^{\widetilde{N}_{\rm slot}}\widehat{\mathbf{H}}'^{\rm H}_{t,d}\widehat{\mathbf{H}}'_{t,d}}{\mathrm{Tr}(\sum_{t=1}^{\widetilde{N}_{\rm slot}}\widehat{\mathbf{H}}'^{\rm H}_{t,d}\widehat{\mathbf{H}}'_{t,d})}
\end{split}
\text{,}
\end{equation}
\begin{equation}\label{R_rx}
\begin{split}
\mathbf{R}^{\rm rx}_{d} = \frac{N_{\rm r}\sum_{t=1}^{\widetilde{N}_{\rm slot}}\widehat{\mathbf{H}}'_{t,d}\widehat{\mathbf{H}}'^{\rm H}_{t,d}}{\mathrm{Tr}(\sum_{t=1}^{\widetilde{N}_{\rm slot}}\widehat{\mathbf{H}}'_{t,d}\widehat{\mathbf{H}}'^{\rm H}_{t,d})}
\end{split}
\text{,}
\end{equation}
respectively, where the trace operation $\mathrm{Tr}(\cdot)$ is performed for normalization. Then the kronecker product is implemented on the transmitting and receiving self-correlation matrices to obtain the joint spatial feature as 
\begin{equation}\label{R_all}
\setlength{\abovedisplayskip}{3pt}
\setlength{\belowdisplayskip}{3pt}
\begin{split}
\mathbf{R}_{d} = \mathbf{R}^{\rm rx}_{d}\otimes\mathbf{R}^{\rm tx}_{d} \in \mathbb{C}^{N_{\rm t}N_{\rm r}\times N_{\rm t}N_{\rm r}}
\end{split}
\text{,}
\end{equation}

It should be noted that the dataset of the target scenario could be very small, and thus it is not sufficient for training autoencoder with superior CSI feedback and to ensure recovery performance, even though it is able to converge quickly based on the initialization $\widehat{\Theta}$ obtained by meta training phase. Therefore, it is necessary to consider a
data augmentation seeded by $\mathcal{H}$ exploiting the  knowledge of statistical features in spatial and delay domain.

The intrinsic knowledge of statistical features of the channel for a specific UE can be completely described by $\hat{p}_d$ and $\mathbf{R}_d, 1\leq d \leq N_{\rm d}$. Therefore, to align with the statistical features of the collected channel samples, the augmented channel for the $d$-th delay $\hat{\mathbf{h}}_d^{\rm aug}$ should satisfy
\begin{equation}\label{R_d}
\begin{split}
\mathbb{E} [ \hat{\mathbf{h}}^{\rm aug}_d (\hat{\mathbf{h}}^{\rm aug}_d)^{\rm H} ]  = \hat{p}_d \mathbf{R}_d
\end{split}
\text{.}
\end{equation}

\subsubsection{Knowledge-driven Target Retraining}
The knowledge-driven target retraining is introduced with data augmentation inspired by (\ref{R_d}).
Firstly, SVD is performed on $\mathbf{R}_d$, i.e.,
\begin{equation}\label{R_svd}
\setlength{\abovedisplayskip}{3pt}
\setlength{\belowdisplayskip}{3pt}
\begin{split}
\mathbf{U}_{d},\mathbf{D}_{d},\mathbf{V}_{d} = \mathrm{svd}(\mathbf{R}_{d})
\end{split}
\text{,}
\end{equation}
where $\mathbf{V}_d = \mathbf{U}_d^{\rm H}$ because of $\mathbf{R}_d = \mathbf{R}_d^{\rm H}$.

Secondly, the augmented channel sample for the $d$-th delay can be generated by conducting
\begin{equation}\label{h_aug}
\begin{split}
\hat{\mathbf{h}}^{\rm aug}_{d} = \sqrt{\hat{p}_{d}}\mathbf{U}_{d}\mathbf{D}_{d}^{\frac{1}{2}}\mathbf{n}
\end{split}
\text{,}
\end{equation}
where the random vector $\mathbf{n} \in \mathbb{C}^{N_{\rm t}N_{\rm r} \times 1}\sim \mathcal{CN}(0,1)$.

Next, $\hat{\mathbf{h}}^{\rm aug}_{d}$ can be reshaped as the channel matrix $\widehat{\mathbf{H}}^{\rm aug}_{d} \in \mathbb{C}^{N_{\rm r} \times N_{\rm t}}$. By concatenating all $N_{\rm d}$ augmented channel matrices, the augmented channel sample can be obtained as
\begin{equation}\label{H_aug}
\begin{split}
\mathbf{H}^{\rm aug} = [\widehat{\mathbf{H}}^{\rm aug}_{1},...,\widehat{\mathbf{H}}^{\rm aug}_{N_{\rm d}}]
\end{split}
\text{,}
\end{equation}
where $\mathbf{H}^{\rm aug} \in \mathbb{C}^{N_{\rm r} \times N_{\rm t} \times N_{\rm d}}$. Then the augmented CSI eigenvector sample $\mathbf{W}^{\rm aug}$ can be finally obtained by implementing (\ref{Hf_1}) to (\ref{Wi}) on $\mathbf{H}^{\rm aug}$.

For each UE, the total $N_{\rm aug}$ channel samples can be provided with $N_{\rm aug}$ randomly generated vectors $\mathbf{n}$. Moreover, for $N_{\rm ue}$ UEs, we can generate totally $N_{\rm ue}N_{\rm aug}$ augmented CSI eigenvector samples that can be used to construct the target task dataset $\mathcal{T}^{\rm aug}_{\rm target}$.

Based on the target task dataset $\mathcal{T}^{\rm aug}_{\rm target}$ and the initialization $\widehat{\Theta}$ obtained in knowledge-driven meta training phase, (\ref{target_retraining_pahse_problem}) can be solved with higher SGCS using a few training steps, i.e.,
\begin{equation}\label{target_retraining_1}
\begin{split}
\Phi = \mathrm{U}^{g}_{\mathcal{T}_{\rm target}}(\widehat{\Theta})
\end{split}
\text{.}
\end{equation}
The proposed algorithm for knowledge-driven target retraining phase can be summarized in Algorithm \ref{alg_2}.

\begin{algorithm}[t]
\caption{Knowledge-driven Target Retraining Phase}
\label{alg_2}
\textbf{Initialization}:$\mathcal{H}$, $N_{\rm aug}$, $g'$\;
Calculate statistical delay power spectrum using (\ref{cal_DPS})\;
\For{$q = 1,\ldots,\widetilde{N}_{\rm ue}$}{
    \For{$d = 1,\ldots,N_{\rm delay}$}{
    Data augmentation using (\ref{R_tx}) to (\ref{H_aug}) and (\ref{Hf_1}) to (\ref{Wi})\;

	}
}
Target retraining using (\ref{target_retraining_1}).
\end{algorithm}

\begin{table}[bp]
\vspace{-0.4cm} 
\setlength{\abovecaptionskip}{-0.01cm}
\centering
\caption{Basic simulation parameters}
\label{tabSystemParameters}
\setlength{\tabcolsep}{4mm}{
\begin{tabular}{|c|c|}
\hline
 Parameter  & Value \\
 \hline
System bandwidth &  10MHz\\
 \hline
Carrier frequency &  3.5GHz\\
 \hline
Subcarrier spacing  &  15KHz\\
 \hline
Subcarriers number $N_{\rm sc}$  &  624\\
 \hline
Subband number $N_\textrm{\rm sb}$  &  13\\
 \hline
Horizontal Tx antenna ports per polarization $N_{\rm h}$ & 8\\
 \hline
Vertical Tx antenna ports per polarization $N_{\rm v}$ & 2\\
 \hline
Tx antenna ports $N_{\rm t}$ & 32 \\
 \hline
Rx antennas $N_{\rm r}$ & 4\\
 \hline
 Meta task enviroment size $T$   &  8000 \\
 \hline
Meta training step size $\epsilon$   &  0.25 \\
 \hline
Step number per task $g$   &  32 \\
 \hline
 Spatial diversity degree $L_{\rm task}$   &  6 \\
 \hline
 Frequency diversity degree $M_{\rm task}$   &  6 \\
 \hline
Spatial diversity scale $\alpha$   &  0.75 \\
 \hline
Frequency diversity scale $\beta$   &  0.75 \\
 \hline
\end{tabular}}
\end{table}

\section{Simulation Results}\label{Simulation Results}
The simulation results are provided in this section. Knowledge-driven scheme in meta training phase (KMeta-*) and target retraing phase (*-KAug) are evaluated, where `None' denotes no knowledge-driven schemes are used. The simulation parameters are listed in Table \ref{tabSystemParameters}. CDL-C \cite{TR0004} channel model with delay spread 300 ns and random distributed UE with speed 300 km/h are utilized as actually collected channels. The training was performed three times with different random seeds, one of which is shown since the results are almost equal. Moreover, the Transformer backbone for CSI feedback \cite{xiao2022ai} with number of feedback bits $B=64$ is implemented in evaluation.

\begin{figure}[t]
\setlength{\abovecaptionskip}{-0.01cm}
\setlength{\belowcaptionskip}{-0.1cm}
\centering
\includegraphics[scale=0.4]{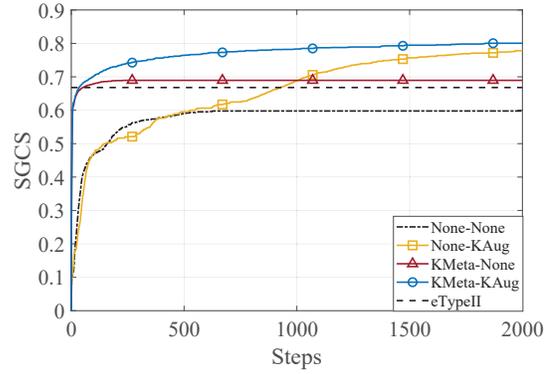}
\caption{Convergence process of target retraining phase with the number of training steps on CDL-C channel ($\widetilde{N}_{\rm ue} = 300$, $\widetilde{N}_{\rm slot} = 10$, $N_{\rm aug} = 100$).}
\label{fig_1}
\vspace{-0.4cm}
\end{figure}

Fig. \ref{fig_1} show the convergence process of target retraining phase with the number of training steps. Note that the vertical axis represents the best achieved SGCS on the test set within the steps. Here we consider the eTypeII codebook and DL-based method without meta training and target augmentation (None-None) as the baselines. Note that since the existing meta-learning methods for CSI require a large amount of multi-scenario real data, and our method lever knowledge for meta-learning, it is unfair to compare our method with them in terms of data cost. In terms of convergence speed, it can be noticed that the proposed KMeta-None require fewer training steps to achieve convergence than None-None. Even on augmented data, KMeta-KAug can also fit more quickly than None-KAug. From the perspective of feedback performance, the knowledge-driven meta training brings higher SGCS since KMeta-None outperforms None-None. Moreover, the methods of *-KAug outperform the methods of *-None, which reveals that the knowledge-driven target retraining phase can further effectively improve the SGCS performance. 

\begin{figure}[t]
\setlength{\abovecaptionskip}{-0.01cm}
\setlength{\belowcaptionskip}{-0.1cm}
\centering
\includegraphics[scale=0.4]{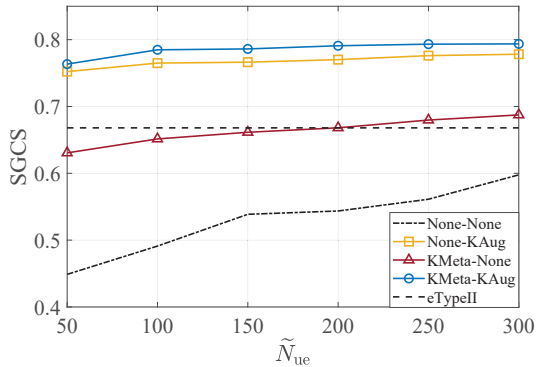}
\caption{Comparison of SGCS for varying number of seeded UEs $\widetilde{N}_{\rm ue}$ on CDL-C channel, fixing number of slots per UE $\widetilde{N}_{\rm slot} = 10$.}
\label{fig_6}
\vspace{-0.4cm}
\end{figure}

\begin{figure}[t]
\setlength{\abovecaptionskip}{-0.01cm}
\setlength{\belowcaptionskip}{-0.1cm}
\centering
\includegraphics[scale=0.4]{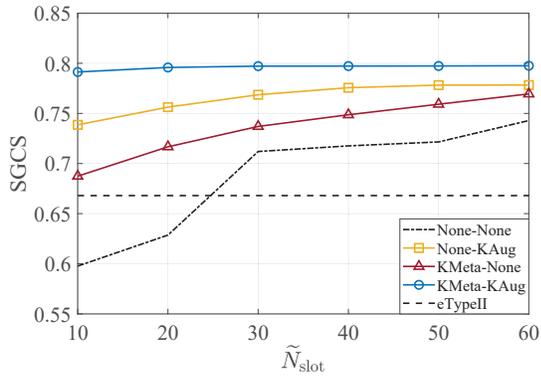}
\caption{Comparison of SGCS for varying number of slots per UE $\widetilde{N}_{\rm slot}$ on CDL-C channel, fixing number of UEs $\widetilde{N}_{\rm ue} = 300$.}
\label{fig_7}
\vspace{-0.4cm}
\end{figure}

In Fig. \ref{fig_6} and Fig. \ref{fig_7} we compare the SGCS performance training 2000 steps on different number of seeded UEs $\widetilde{N}_{\rm ue}$ and slots $\widetilde{N}_{\rm slot}$, respectively. It is observed that the proposed knowledge-driven method of KMeta-KAug outperforms traditional eTypeII codebook and basic DL-based method None-None. Specifically, the performance gaps between the methods KMeta-* and None-* can respectively demonstrate the gain provided by proposed knowledge-driven meta training. The gaps between *-KAug and *-None respectively initimate the gain obtained from proposed knowledge-driven target retraining. Moervoer, in Fig. \ref{fig_7}, the performance of None-KAug improves as the number of slots $\widetilde{N}_{\rm slot}$ increased, while the performance of KMeta-KAug stays almost unchanged, which implies that the proposed knowledge-driven target retraining requires fewer slots to achieve the performance ceiling when it is enhanced by proposed knowledge-driven meta training.

\begin{figure}[t]
\setlength{\abovecaptionskip}{-0.01cm}
\setlength{\belowcaptionskip}{-0.1cm}
\centering
\includegraphics[scale=0.4]{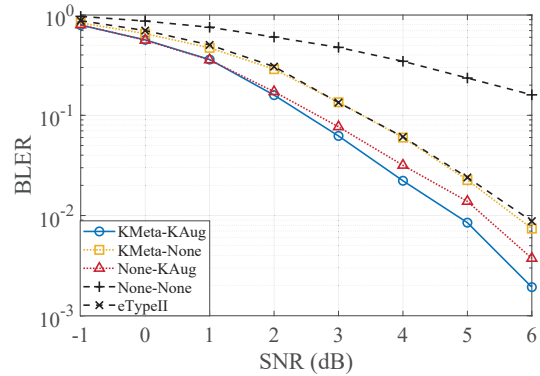}
\caption{Link-level BLER performance comparison trained on CDL-C300 for different solutions ($\widetilde{N}_{\rm ue} = 300$, $\widetilde{N}_{\rm slot} = 10$).}
\label{BLER}
\vspace{-0.4cm}
\end{figure}

\begin{table}[b]
\vspace{-0.5cm} 
\setlength{\abovecaptionskip}{-0.01cm}
\centering
\caption{Comparison of different augment schemes}
\label{different_aug_scheme}
\setlength{\tabcolsep}{15mm}{
\begin{tabular}{|c|c|}
\hline
Scheme          & SGCS   \\ \hline
None            & 0.5977 \\ \hline
Noise Injection & 0.6189 \\ \hline
Flipping        & 0.6171 \\ \hline
Cyclic Shift    & 0.6404 \\ \hline
Random Shift    & 0.6225 \\ \hline
Rotation        & 0.6178 \\ \hline
Proposed        & \textbf{0.7930} \\ \hline
\end{tabular}}
\begin{flushleft}
Note 1: $\widetilde{N}_{\rm ue} = 300$ and $\widetilde{N}_{\rm slot} = 10$

Note 2: All methods augments to 30k samples, except the flipping which can only augment to 6k samples due to method limitation. 
\end{flushleft}
\end{table}

Table \ref{different_aug_scheme} illustrates the SGCS performance of  the proposed method and the existing data augmentation methods \cite{xiao2022ai} for DL-based CSI feedback including noise injection, flipping, cyclic shift, random shift and rotation. It is observed that the proposed method can obtain 0.1953 SGCS performance gain in comparison to none augmantation. Specifically, the performance gap between proposed method and other competitors is at least 0.1526, which demonstrates that exploiting communication knowledge effectively bring performance gain.

The link-level block error rate (BLER) performance is presented in Fig. \ref{BLER}, where omnidirectional and directional antennas is deployed at UE and BS, respectively. The gap between KMeta-None and None-None proves the performance gain of knowledge-driven meta training phase. The gap between None-KAug and None-None proves the performance gain of knowledge-driven target retraining phase. Since the method of KMeta-Aug outperforms other competitors in terms of BLER, the advantages and application potential of proposed knowledge-driven approach are well demonstrated.

\section{Conclusion}\label{Conclusion}
In this paper, we propose a knowledge-driven meta-learning method for CSI feedback, where the meta task environment for meta training is constructed based on the intrinsic knowledge of spatial-frequency feature of CSI eigenvector. Initialized by the knowledge-driven meta training phase, the DL model is capable of achieving rapid convergence by retraining on the target task dataset, which is augmented from only a few actually collected seeded data with the assistance of the knowledge of statistical feature of wireless channels. Simulation results demonstrate the superiority of the approach from the perspective of feedback performance and convergence speed.

\vspace{-0.2cm}
\bibliographystyle{IEEEtran}
\bibliography{IEEEabrv,ref}

\end{document}